# Metamaterial insertions for resistive-wall beam-coupling impedance reduction

A. Danisi, C. Zannini, R. Losito, A. Masi

CERN, Geneva, Switzerland

Resistive-wall impedance usually constitutes a significant percentage of the total beam-coupling impedance budget of an accelerator. Reduction techniques often entail high electrical-conductivity coatings. This paper investigates the use of negative-permittivity or negative-permeability materials for sensibly reducing or theoretically nearly cancelling resistive-wall impedance. The proposed approach is developed by means of an equivalent transmission-line model. The effectiveness of such materials is benchmarked for the negative permeability case with experimental measurements in two frequency bandwidths. This first-stage study opens the possibility of considering metamaterials for impedance mitigation.

## I. INTRODUCTION

The resistive-wall term of beam-coupling impedance is essentially due to the finite electrical conductivity of the beam pipe walls [1]. Many countermeasures are often used to mitigate this, as conductive coating or ceramic inserts [2]. In particular, the choice of coating materials is crucial.

Metamaterials or, more in detail, composite materials with negative values of either relative permittivity or relative permeability have been intensively studied in the last decades [3]-[4], in the framework of RF cloaking [3], [5], as well as in waveguides [3]. Some publications also addressed the application of metamaterials in particle accelerators [6].

Concerning metamaterials insertions for beam-coupling impedance mitigation, their effect has been first addressed in [7]. This paper extends such work, by evaluating the impact of metamaterial insertions (in the form of layers applied to the beam pipe wall) for the reduction of resistive-wall beam-coupling impedance both from the theoretical and experimental viewpoints.

Section II introduces the properties of metamaterials and their fabrication. Section III describes the theoretical analysis using a transmission-line model, whereas Section IV gives details on the experimental measurements which have been performed with simple negative-permeability metamaterials.

## II. PROPERTIES OF METAMATERIALS

The behaviour of electromagnetic waves in media with negative value of permittivity and/or permeability has been addressed with a number of works in literature [8]-[10]. In particular, in a composite material, made of inclusions embedded in a host medium (surface or volume), the electromagnetic waves interact with the inclusions according to numerous variables (e.g. conductivity, permittivity of host medium, dimensions, density, alignment) and therefore changes the equivalent global constitutive parameter of the material, i.e. its macroscopic permittivity or permeability [3], [10]. It is evident that such values will basically be a function of the scale of the wave-medium interaction, i.e. the ratio between the wavelength and the dimensions of the inclusions [3].

In other words, according to the size of the inclusions and the constitutive parameters of the host medium, metamaterials can be designed to show the desired macroscopic value of permittivity (or permeability) in a limited frequency range [3], [11].

In order to model the frequency dependence of the macroscopic ε or μ, the two-time-derivative Lorentz metamaterial (2TDLM) model [3] is often used. It describes the relation between the electric polarization (or magnetization) and the electric (or magnetic) field, satisfying both causality and the generalized Kramers-Kronig relations [3], [12].

Experiments and theoretical analyses have shown [3], [4] that when inserted in standard waveguides (e.g. rectangular), metamaterials exhibit different behaviour according to the mutual sign of their constitutive parameter. For example, metamaterials with only one negative constitutive parameter (MNG for negative-valued permeability, ENG for negative-valued permittivity) are associated to evanescent waves [4], and therefore can give rise to stopbands even above the typical cutoff frequency of the waveguide [3]. On turn, double-negative metamaterials (DNG) give rise to backward waves, in which the power direction is opposite as for conventional materials [3], [13].

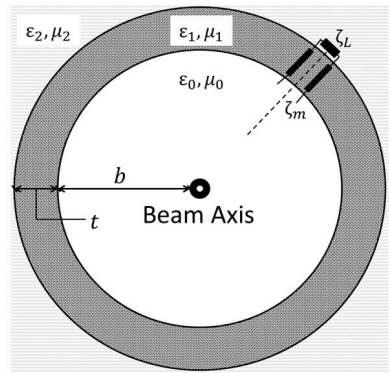

Figure 1: Transverse geometry of the problem. The metamaterial layer is put between the vacuum and the beam pipe (which extends to infinity).

Metamaterials are often synthesized by mixing a matrix of conductive inclusions in an isotropic host medium. A simple way to achieve a single-negative metamaterial (in particular, a MNG material) is to build a split-ring resonator (SRR) [3], [9]. This structure is made of two conductive copper rings which can be easily printed on a substrate (e.g. fiberglass) and then repeated with a periodic structure. According to the inclusions' dimensions and electrical permittivity of the substrate, this composite material exhibits MNG behaviour around a specified resonance frequency [9].

## III. THEORETICAL ANALYSIS

*Transmission-line equivalence*

The maximum acceptable impedance of a particle accelerator (i.e. the "impedance budget") is following a decreasing trend for modern machines (e.g. HL-LHC [14]-[15]). On the other hand, many existing and future machines present very short bunches, leading to a much wider frequency range of interest [15].

The theoretical analysis presented in this section reduces the geometry of the problem (depicted in Fig. 1) to an equivalent transmission-line problem. As explained in [16], the resistive wall impedance of such a geometry can be easily calculated by means of an impedance transportation of the wall surface impedance along the different equivalent transmission lines formed by the coating layers. In the case of Fig. 1, just one layer is present, and it is a MNG or an ENG material.

The transverse impedance is computed from the longitudinal one [16].

The condition of validity of the transmission line approach is that the equivalent surface impedance (computed transporting the layer impedances to the first interface) is constant with the incident wave direction [7]. This condition is satisfied when the permittivity and permeability of the first layer are such that [7]

$$|\varepsilon_1 \mu_1| \gg \varepsilon_0 \mu_0 \quad (1)$$

where $\varepsilon_0$ and $\mu_0$ denote vacuum permittivity and permeability. It has to be noted that (1) can still be applied for $\varepsilon_1 < 0$ or $\mu_1 < 0$. As discussed in [16], for ultra-relativistic beams the transmission-line approach gives satisfactory results.

Following this approach, the longitudinal resistive-wall impedance can be written as [16]

$$Z_\parallel = \frac{l \cdot \zeta_T}{2\pi b} \quad (2)$$

where $\zeta_T$ is the impedance seen at the vacuum-layer interface, $b$ the radius at this interface and $l$ is the length of the pipe sector (Fig. 1).

If

$$\zeta_S = \frac{1+j}{\sigma \delta} \quad (3)$$

is the wall surface impedance ($\sigma$ and $\delta$ are the wall conductivity and penetration depth, respectively), and taking into consideration Fig. 1 with $\varepsilon_1 > 0$, $\mu_1 < 0$ (MNG layer), the real and imaginary parts of $\zeta_T$ can be written as

$$Re\{\zeta_T\} = \zeta_R \cdot \frac{A[1 - (\tanh k_R t)^2]}{(A - \tanh k_R t)^2 + (\tanh k_R t)^2}$$

$$Im\{\zeta_T\} = \zeta_R \cdot \frac{A(1 - A \tanh k_R t) - (\tanh k_R t)(2 - At)}{(A - \tanh k_R t)^2 + (\tanh k_R t)^2} \quad (4)$$

where

$$A = \sigma \delta \zeta_R$$

$$\zeta_R = \sqrt{\frac{|\mu_1|}{|\varepsilon_1|}} \quad (5)$$

$$k_R = \omega \sqrt{|\varepsilon_1| \cdot |\mu_1|}$$

These results have been obtained by simply transporting $\zeta_S$ along an MNG transmission line, choosing the correct determination of the square roots for impedance and wave number [4]. Eq. (4) are consistent, since

$$\lim_{\sigma \to \infty} Re\{\zeta_T\} = 0$$

$$\lim_{\sigma \to \infty} Im\{\zeta_T\} = -j\zeta_R \tanh(k_R t) \quad (6)$$

which corresponds to the impedance transportation of a short circuit (i.e. perfect electrical conductor) along the MNG line. The same procedure can be followed for ENG layers, leading to similar and complementary results.

The development of the impedance transportation, and successively the computation of the resistive-wall impedance, is performed through TL-Wall, a MATLAB-based code written for the purpose of evaluating the impedance with different coating layers [16] adopting a transmission-line equivalence.

*Metamaterial Impact*

Fig. 1 shows the geometry of the electromagnetic problem. The metamaterial insertion covers the internal side of the beam pipe by a thickness $t$. The thickness of the beam pipe wall, modelled as a conductive medium with $\varepsilon_2 = \varepsilon_0$ and $\mu_2 = \mu_0$, is assumed to be infinite. This approximation holds for small skin depths (compared to the pipe wall thickness).

Fig. 2 and Fig. 3 show the calculated longitudinal and transverse (in this case only dipolar) impedances through TL-Wall when an insertion layer of $t = 10$ mm with a negative permittivity or a negative permeability is alternatively considered. The parameters used for the calculation are listed in Table 1. For this first theoretical analysis, the constitutive parameters have been assumed frequency-independent.

In the equivalent transmission-line model, the insertion of a layer entails an additional line sector, to be modelled with its characteristic impedance and wave number. In the case of a metamaterial, the characteristic impedance is imaginary. In fact, the ENG and the MNG material lines are described by inductive and capacitive characteristic impedance, respectively. This means that, as observed for metamaterials in general [3], [4], the electromagnetic wave in these sectors is evanescent. Therefore, the impedance transportation along such a line leads to an intrinsic alteration of the resistive-wall impedance.

As a matter of fact, Fig. 2 shows that both ENG and MNG layers determine a major decrease of the real part of longitudinal impedance above a characteristic frequency (in this case, about 300 MHz for ENG and 5 GHz for MNG insertions), which in principle depends both on constitutive parameters and on layer thickness. Nevertheless, the impact is significant, since the real part decreases by several orders of magnitude (e.g. more than 10, for ENG insertions). The influence on the imaginary part is instead opposite: whereas the MNG insertion translates the imaginary part from inductive to capacitive (the negative part is not displayed in logarithmic scale), the ENG layer increases it up to a constant above the previously-defined characteristic frequency.

On the transverse plane, the effect is also evident. The MNG insertion increases the real part and decreases the imaginary one, down to capacitive (i.e. negative) values.

Instead, the ENG layer significantly decreases the real part and increases the imaginary one. In both cases though, the imaginary parts are approaching zero above their characteristic frequencies.

Other than for this first-approach case, the impact of the metamaterial layer has been clearly observed even with much smaller values of the thickness $t$. Overall, the observed results demonstrate a remarkable influence on the resistive-wall beam-coupling impedance, which can lead to the individuation of theoretical design rules for impedance mitigation, exploiting the different degrees of freedom which emerged from this analysis: the type of material (ENG or MNG), its values of constitutive parameters, its thickness and its length. As a consequence, a proper engineering of such insertions can be performed, with the aim of substantially reducing the resistive-wall impedance of a beam line.

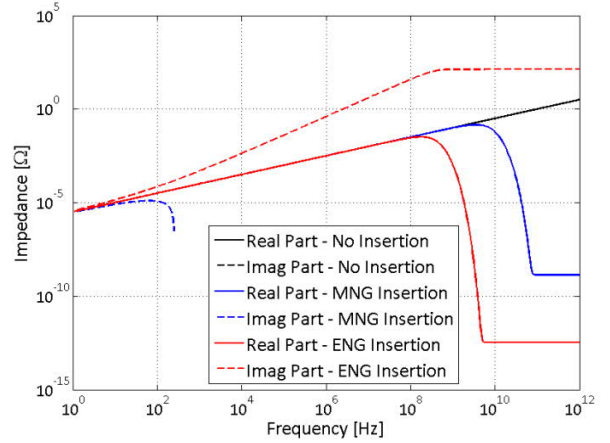

Figure 2: Effect of metamaterial insertions on real and imaginary part of longitudinal impedance.

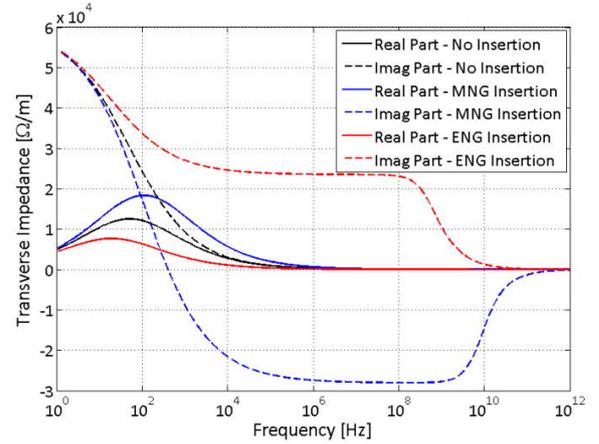

Figure 3: Effect of metamaterial insertions on real and imaginary part of transverse impedance.

As an example, for the considered 1-m-long cylindrical pipe, Fig. 4 shows the results when a simple linear rule has been adopted to distribute ENG and MNG layers along the length, in order to integrate the effect of both.

The results show that the longitudinal impedance is substantially decreased, whilst in the transverse plane, a slight increase of the real part is observed. The imaginary part is also decreased, down to negative values (but smaller, in absolute value, than the ones obtainable with 100 % of MNG material, as testified in Fig. 3). This kind of arrangement would, for example, suit a need for lowering the longitudinal real part, with some margin on the transverse plane. On the other hand, the introduction of capacitive impedance on this plane may also be useful to minimize the total transverse impedance.

This example of metamaterial insertion engineering has taken into account just one degree of freedom. A wider and more complex optimization can be thought *ad-hoc* considering all design parameters.

Table 1: Parameters of the Electromagnetic Problem.

| Parameter | Value |
| --- | --- |
| $b$ [mm] | 31.5 |
| $t$ [mm] | 10.0 |
| Length of the line [m] | 1 |
| $\varepsilon_{r,1} = \varepsilon' + j\varepsilon''$ $\mu_{r,1} = \mu' + j\mu''$ (for ENG insertions) | $\varepsilon_{r,1} = -200 + j \cdot 10^{-12}$ $\mu_{r,1} = 1$ |
| $\varepsilon_{r,1} = \varepsilon' + j\varepsilon''$ $\mu_{r,1} = \mu' + j\mu''$ (for MNG insertions) | $\varepsilon_{r,1} = 1$ $\mu_{r,1} = -0.5 + j \cdot 10^{-12}$ |
| $\sigma_{el}$ (conductivity of beam pipe wall) [S/m] | $10^7$ |

*Re-Wall Benchmarking*

In order to assess the reliability of the results obtained with TL-wall, the same metamaterial layers have been submitted to Re-Wall, a different code which computes the full 2D field solutions and matches them at the interfaces [17]. The results, more deeply discussed in [7], show very good agreement.

## IV. EXPERIMENTAL VALIDATION

As discussed in Section II, metamaterials can be manufactured by using inclusions in a host medium. However, it is impossible to synthesize a material with MNG or ENG properties along the whole frequency spectrum.

In this section, experimental measurements are shown using split-ring resonators (SRRs) as sample metamaterial. The aim of such measurements is to have a first proof of concept, highlighting the decrease of the resistive-wall impedance due to the metamaterial insertions, through quality factor measurements.

*Experimental Setup and Procedure*

The configuration and dimensions of the SRRs used for such measurements are shown in Fig. 5. The tracks are made of copper and they have been printed on a G-10 fiberglass substrate.

The measurements have been performed by evaluating the quality factor Q of several resonances in a cavity. The cavity has been obtained using a straight section of a rectangular waveguide, enclosing it between two metallic plates and inserting a tiny antenna on one side, in order to excite the modes. In particular, two different cavities have been synthesized, using WR284 and WR187 straight waveguide sections, in order to cover a wider bandwidth.

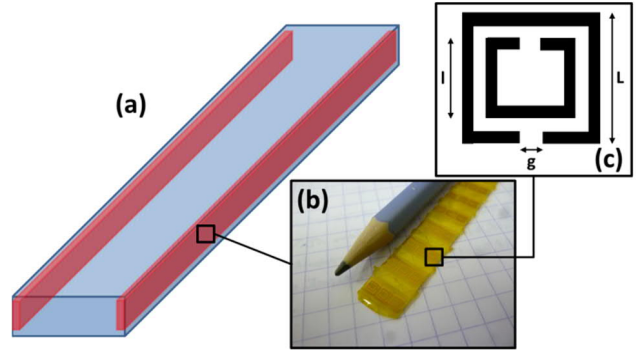

Figure 5: (a) Illustration of the rectangular cavity with (in red) the metamaterial strips; (b) picture of the SRR strips; (c) particular showing the split-ring resonator geometry, with L = 5.4 mm, l = 3.4 mm, g = 1 mm.

The $S_{11}$ scattering parameter has been acquired through a Vectorial Network Analyzer (VNA) in the two different bandwidths. The same acquisition has been repeated placing two SRR stripes on the cavity walls, as depicted in Fig. 5.

The resonance frequencies on the $S_{11}$ spectra have been individuated and the corresponding Q factors have been measured using the VNA and post-processing techniques.

The acquired Q factors are necessarily affected by the presence of the antenna (i.e. they are 'loaded' Qs [18]) and therefore they do not directly correspond to the real Q factors of the cavity (i.e. the 'unloaded' Qs). To unload the measured Q factor for each resonance, the following procedure has been considered:

1) acquire the resonance spectrum with a relatively small span (e.g. 10 MHz) around the peak;
2) verify, through a $S_{11}$ plot on the Smith chart, if the resonance is being measured in under-coupling or over-coupling conditions of the antenna [19];
3) calculate the Voltage Standing Wave Ratio (VSWR) and determine the coupling factor β from its minimum [19], according to the coupling condition verified in 2);
4) apply the correction on the measured loaded Q to obtain the unloaded quality factors.

*Measurements Results*

The unloaded quality factor of a resonance associated to an empty waveguide section is related solely to the losses on the conductive walls. Such losses are due to the surface impedance of the walls, which depends on the material conductivity and frequency. The quality factor is inversely proportional to the surface resistance [19].

If the waveguide is regarded as a beam pipe, the surface impedance gives rise to a resistive-wall component of the beam-coupling impedance, as explained in Section III. In particular, the resistive-wall impedance is proportional to the surface impedance.

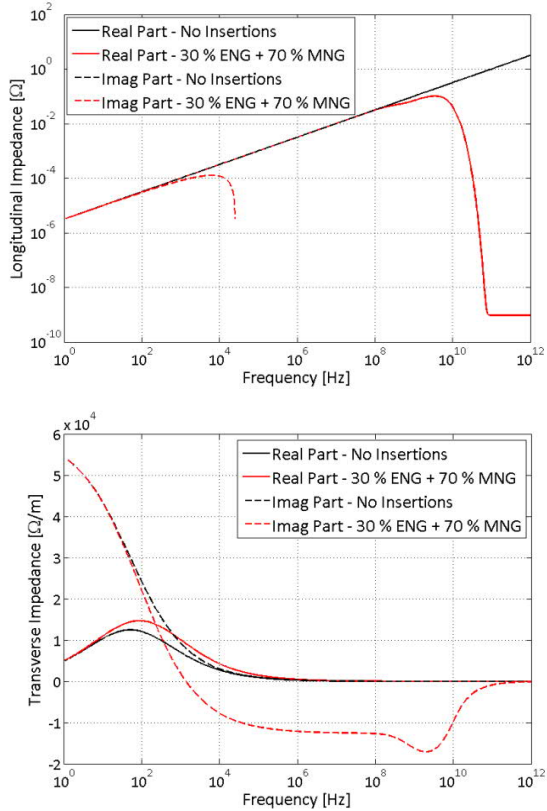

Figure 4: Longitudinal and transverse impedances of a cylindrical beam pipe as of Table 1, loaded with 30 % ENG and 70 % MNG layers along its length.

Therefore, to interpret the results coming from the measurements, one should note that an increase of the unloaded quality factor testifies a proportional decrease of the resistive-wall impedance (caused by a decrease of the surface impedance); conversely, a decrease of the unloaded Q means the resistive-wall impedance is proportionally increasing at that frequency.

The measurement results are shown in Fig. 6 for both bandwidths (in S-band and C-band, respectively). First of all, it is noticeable in both bands that the presence of the SRR insertions shifts the resonance frequencies downwards. The results in S-band show that at about 2.9 GHz the measured unloaded Q is significantly higher in the case of metamaterial presence, meaning a decrease of an order of magnitude of the surface impedance, i.e. of the resistive-wall impedance. The other peaks instead show little or no variation when the SRRs are put in the waveguide.

The results in C band also testify a decrease of the resistive-wall impedance at about 4.1 GHz and 4.55 GHz, since their quality factor is increased by 30 % and 50 % respectively. Instead, the last point, at 5.1 GHz, reports a worsening of the unloaded Q.

From these measurements, the following observations are then to be pointed out
  i.  Two frequency regions are identified, i.e. around 2.9

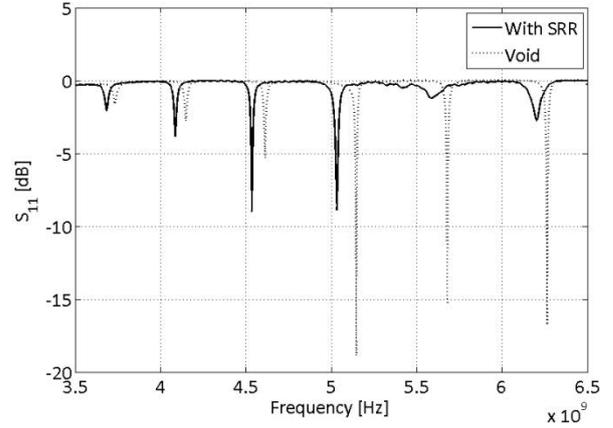

Figure 7: $S_{11}$ scattering parameter measured in C band.

GHz and from 4.1 GHz to 4.55 GHz, where there actually is a benefit in reducing the resistive-wall impedance thanks to SRR insertions.
  ii. Fig. 7 shows the full-span acquisition in C band. It is interesting to note that above 5.5 GHz the SRR influence is more in damping the modes, rather than improving the unloaded Q. However, it is possible that with the presence of the SRRs, the cutoff frequency of the first higher-order mode is lowered [3], and therefore in this frequency region it could already propagate, affecting the Qs. For this reason, the analysis should consider frequencies below 5.5 GHz. For the same reason, the analysis in S band is limited to 3.8 GHz.
  iii. Given that the frequency regions where the SRRs show their effectiveness depend on their size and host material properties, a proper engineering of such parameters can bring to a wider bandwidth or to a proper tuning of such materials over the desired frequency range.

## VI. CONCLUSIONS AND OUTLOOK

The effect of metamaterial insertions on beam-coupling impedance is studied theoretically by means of a transmission-line model and benchmarked with a 2D full-field code. Metamaterials showed to impact the resistive-wall impedance both in the longitudinal and transverse planes. As a proof of principle, experimental measurements are performed with sample SRR metamaterials. The results show a significant decrease of resistive-wall beam-coupling impedance around 2 frequency regions.

More complex studies, regarding the design of proper metamaterials for frequency tuning, as well as introducing a frequency-dependent permeability in the theoretical mode, are foreseen. Further engineering of metamaterial properties (dimensions, shape etc.) as well as a suitability study for accelerator environments (ultra-high vacuum, radiation etc.) are also in progress.

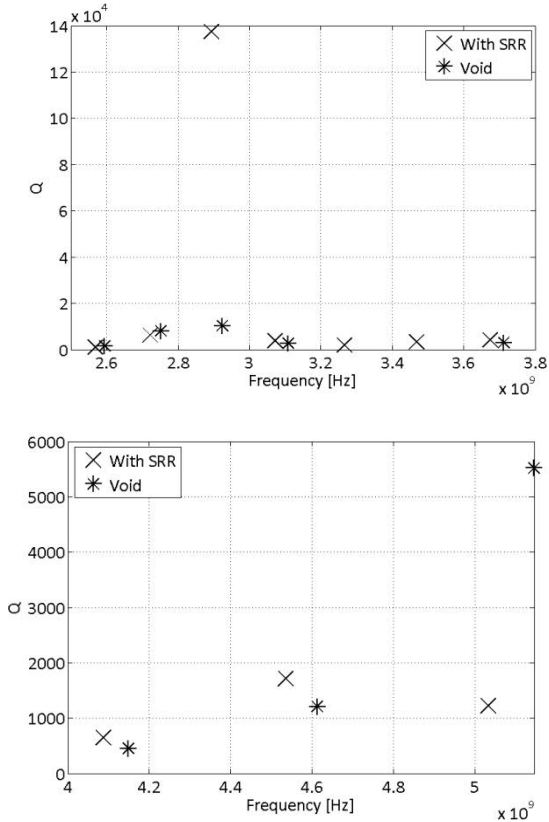

Figure 6: Measured unloaded quality factors with and without SRR insertions for (Top) S band and (Bottom) C band.